\documentclass[newstyle,twocolumn,proceedings]{rmaa}

\usepackage{rmaacite}

\renewcommand{\P}[1]{%
\ifnum#1=1\hbox{OW~168--326E}\fi
\ifnum#1=2\hbox{OW~167--317}\fi
\ifnum#1=3\hbox{OW~163--317}\fi
\ifnum#1=5\hbox{OW~158--323}\fi
\ifnum#1=0\hbox{OW~171--334}\fi}

\title{The Temperature Structure in Ionized Nebulae and the Chemical Evolution
of Galaxies}
\author{Manuel Peimbert and Antonio Peimbert
  \affil{Instituto de Astronom\'{\i}a, UNAM} }

\fulladdresses{
\item M. Peimbert and A. Peimbert: Instituto de Astronom\'{\i}a,
  UNAM, Apartado Postal 70-264, M\'exico, D.F., 04510, M\'exico
  (peimbert,antonio@astroscu.unam.mx).}

\shortauthor{Peimbert \& Peimbert}
\shorttitle{Temperature Structure in Ionized Nebulae}


\abstract{A few results that indicate the presence of temperature variations in
gaseous nebulae are reviewed. The evidence is based on: a) temperatures derived
from different methods, and b) on comparisons of abundances predicted by models
of galactic chemical evolution with abundances derived from observations. }
\resumen{Discutimos algunos resultados que indican la presencia de variaciones
de temperatura en nebulosas gaseosas. La evidencia se basa en: a) temperaturas
obtenidas a partir de m\'etodos diferentes, y b) en la comparaci\'on de
abundancias predichas por modelos de evoluci\'on qu\'{\i}mica de galaxias con
abundancias determinadas a partir de observaciones. }

\listofauthors{M.~Peimbert \& A.~Peimbert}
\indexauthor{Peimbert, M.}
\indexauthor{Peimbert, A.}

\begin{document}

\maketitle

\section{Overview}

To constrain models of stellar evolution and of chemical evolution of galaxies
it is necessary to derive accurate abundances from gaseous nebulae.  Inversely,
it might be possible to constrain the assumptions made to derive the abundances
from gaseous nebulae based on robust models of stellar evolution and of
chemical evolution of galaxies.

The determination of the chemical abundances in gaseous nebulae depends
strongly on their temperature structure.  By adopting the temperature derived
from the 4363/5007 [\ion{O}{3}] ratio and assuming that the temperature is
constant it can be shown that in the presence of temperature variations the
abundances are overestimated (\pcite{pei67}; \pcite{pei69}). Typical overestimates of the O/H
abundances are in the 1.4 to 3.0 range, with extreme cases reaching
overabundances of an order of magnitude.

We compare the predicted abundances based on models of galactic chemical
evolution with the abundances derived under the assumption of: a) a constant
temperature distribution and b) the presence of temperature fluctuations.  We
conclude that significant temperature variations are present in gaseous
nebulae.

Recent reviews on the temperature structure of gaseous nebulae are those of
\scite{est02z}, \scite{sta02} and \scite{pei02x}. Recent reviews on the
abundances of galactic and extragalactic \ion{H}{2} regions are those by
\scite{ski98}, Peimbert (1999, 2002), \scite{gar99} and Peimbert, Carigi, \&
Peimbert (2001).

\section{Temperature Structure}

We can characterize the temperature structure of a gaseous nebula by two
parameters: the average temperature, $T_0$, and the mean square temperature
fluctuation, $t^2$, given by

\begin{equation}
T_0 (N_e, N_i) = \frac{\int T_e({\bf r}) N_e({\bf r}) N_i({\bf r}) dV}
{\int N_e({\bf r}) N_i({\bf r}) dV},
\end{equation}

\noindent and

\begin{equation}
t^2 = \frac{\int (T_e - T_0)^2 N_e N_i dV}{T_0^2 \int N_e N_i dV},
\end{equation}

\noindent respectively, where $N_e$ and $N_i$ are the electron and the ion
densities of the observed emission line and $V$ is the observed volume
\cite{pei67}.

For a nebula where all the O is twice ionized we can derive $T_0$ and $t^2$
from the ratio of the [\ion{O}{3}] $\lambda\lambda$ 4363, 5007 lines,
$T_e(4363/5007)$, and the temperature derived from the ratio of the Balmer
continuum to $I({\rm H}\beta)$, $T_e({\rm Bac}/{\rm H}\beta)$, that are given
by

\begin{equation}
T_e(4363/5007) = T_0 \left[ 1 + {\frac{1}{2}}\left({\frac{90800}{T_0}} - 3
\right) t^2\right],
\end{equation}

\noindent and

\begin{equation}
T_e({\rm Bac}/{\rm H}\beta) = T_0 (1 - 1.70 t^2),
\end{equation}

\noindent respectively.

It is also possible to use the intensity ratio of a collisionally excited line
of an element $p + 1$ times ionized to a recombination line of the same element
$p$ times ionized, this ratio is independent of the element abundance and
depends only on the electron temperature. In this review we will adopt the view
that the \ion{C}{2} 4267 and the \ion{O}{2} permitted lines of multiplet 1, are
produced by recombination only and consequently that other mechanisms like
radiative transfer, collisions, and fluorescence do not affect their
intensities.

When oxygen is present both in the form of O$^{++}$ and O$^+$, then there will
be different $T_0$ and $t^2$ values for the O$^{++}$ and the O$^+$ zones. A
full analytic treatment to second order of this case can be found in Peimbert,
Peimbert, \& Luridiana (2002).

The comparison of $T_e(4363/5007)$ with $T_e({\rm Bac}/{\rm H}\beta)$ in
planetary nebulae clearly indicates that $t^2 > 0.000$ and that the temperature
structure needs to be taken into account to determine the abundances
(e.g. \pcite{pei71}; \pcite{liu01}; \pcite{liu02} and references therein).

In \ion{H}{2} regions an important fraction of the volume is in the form of
O$^+$, and $T_e({\rm Bac}/{\rm H}\beta)$ has to be compared with a temperature
derived from a weighted average of collisionally excited lines in the O$^+$ and
O$^{++}$ zones. From photoionization models for \ion{H}{2} regions by
\scite{sta90} the O$^+$ zones are expected to be hotter than the O$^{++}$ zones
for objects with $T_0 <$ 12400 K, which is the case for galactic \ion{H}{2}
regions of the solar vicinity, prediction that has been confirmed by
observations (Esteban et al. 1998, 1999a, 1999b). From a comparison of
$T_e({\rm Bac}/{\rm H}\beta)$ with a weighted average of colisionally excited
lines in the O$^+$ and O$^{++}$ zones it is also found that $t^2 > 0.000$.

A large number of papers have been presented in favor and against the idea that
the recombination lines of heavy elements can be used to derive the chemical
abundances of gaseous nebulae. A strong argument in favor that they are
representative of the abundances of gaseous nebulae is given by the work of
\scite{liu01} where they find a very strong correlation between the O$^{++}$
abundance ratio derived from the \ion{O}{2} to [\ion{O}{3}] lines and the
$T_e(4363/5007) - T_e({\rm Bac}/{\rm H}\beta)$ values, this correlation has
been studied further by \scite{pei02z} who find that it can be explained
quantitatively as being due to temperature fluctuations.


From observations it has been found that $0.01 \leq t^2 \leq 0.09$, with
typical values around $0.04$; while from photoionization models of chemically
and density homogeneous nebulae it has been found that $0.000 \leq t^2 \leq
0.020$, with typical values around $0.005$. An explanation for the differences
in the $t^2$ values predicted by photoionization models and those found by
observations has to be sought (see \pcite{tor02}).

\section{O/H in the Orion Nebula and the Sun}

The determination of the O/H value is paramount for the study of chemical
evolution for at least three reasons: a) O is the most abundant heavy element,
it comprises about 50\% of the heavy elements, b) it is easily observed in the
main ionization stages of \ion{H}{2} regions and planetary nebulae, c) its
atomic physics parameters have been determined with high precision, and d)
practically all of it has been produced by supernovae of Type II, therefore in
chemical abundance models can be treated under the instant recycling
approximation.

A decade ago the O/H solar abundance in the literature was 0.44dex higher than
in the Orion nebula (see Table~\ref{to/h}), a result of great concern because
from galactic chemical evolution models the Orion nebula O/H value is expected
to be about 0.05dex higher than the solar value, if the Sun was formed at the
same galactocentric distance than the Orion nebula (e. g. \pcite{car96}). Note
that the Orion nebula is about 0.4kpc farther away than the Sun from the
galactic center, and from an O/H gradient of -0.05dex~kpc$^{-1}$, a correction of
about 0.02dex has to be made if the Sun were in a circular orbit; nevertheless
this correction might be compensated by the ellipticity of the solar orbit that
probably implies that the Sun was formed at a larger galactocentric distance,
about 0.5 to 1kpc farther away from the galactic center than its present
position.

\begin{table}[htb]
\begin{center}
\caption{Orion and Solar Oxygen Abundances}
\label{to/h}
\begin{tabular}{l@{\hspace{12pt}}c@{\hspace{12pt}}c}
\hline
& Orion Nebula$^a$ & Sun$^a$ \\
\hline 
1980's & 8.49$^b$ ($t^2=0.000$, gas) & 8.93$^d$ \\
1998   & 8.64$^c$ ($t^2=0.024$, gas) & 8.83$^e$ \\[1.3ex]
now    & 8.72$^c$ ($t^2=0.024$, gas + dust) & 8.73$^f$ \\
&& 8.69$\pm$0.05$^g$ \\
\hline
\multicolumn{3}{l}{$^a$ Given in log O/H + 12.}\\
\multicolumn{3}{l}{$^b$ \pcite{sha83}; \pcite{ost92};}\\
\multicolumn{3}{l}{\hspace{12pt} \pcite{rub93}; \pcite{deh00}.}\\
\multicolumn{3}{l}{$^c$ \pcite{est98}.}\\
\multicolumn{3}{l}{$^d$ \pcite{gre89}.}\\
\multicolumn{3}{l}{$^e$ \pcite{gre98}.}\\
\multicolumn{3}{l}{$^f$ \pcite{hol01}.}\\
\multicolumn{3}{l}{$^g$ \pcite{all01}.}\\
\end{tabular}
\end{center}
\end{table}

At present the best Orion nebula O/H value in the literature is 0.02dex higher
than the solar value, the difference between models and observations is well
within the observational errors. The change in the last ten years is due to two
recent results for Orion and two for the Sun: a) the 0.15dex increase in the
O/H value derived from recombination lines (which implies a $t^2 = 0.024$)
relative to that derived from forbidden lines assuming that $t^2 = 0.000$, b)
the increase in the Orion O/H value of 0.08dex due to the fraction of oxygen
embedded in dust grains, and c) the decrease of 0.23dex due to two new solar
determinations (see Table~\ref{to/h}).

\section{Abundance Gradients in Our Galaxy} 

In Table 2 we present average gradients for gaseous nebulae based on results by
many authors (see reviews by \pcite{est95}, \pcite{mac97}, \pcite{pei99} and
references therein), most of the gradients have been determined from
colisionally excited lines. The values come from \ion{H}{2} regions of the
solar vicinity and for O, Ne and S were derived from more than 40 objects. The
planetary nebulae values by \scite{mac99} were derived from a data set of 130
planetary nebulae of Type II. \scite{rol00} from approximately 80 B stars of
the solar vicinity derived an O gradient of $0.061 \pm 0.005$ dex kpc$^{-1}$ ,
and similar gradients for Mg, Al and Si.  Chemical evolution models of the
Galaxy predict very similar gradients for the elements included in Table 2
(e.g. \pcite{all98}; \pcite{chi01}; \pcite{ali02}), since most of their
production is due to massive stars that explode as supernovae of Type
II. Therefore for the elements in Table 2 we expect gradients of about ---
$0.05$ to $-0.06$ dex kpc$^{-1}$.

\begin{table*}[htb]
\begin{center}
\caption{Galactic Abundance Gradients}
\label{GAG}
\begin{tabular}
{c@{\hspace{36pt}}c@{\hspace{18pt}}c@{\hspace{18pt}}c@{\hspace{18pt}}c}
\hline
Gradient & Planetary & \ion{H}{2}
& \multicolumn{2}{c}{(M17$^b$, M8$^c$, Orion$^d$)}\\
\cline{4-5}\\[-9pt]
(dex kpc$^{-1}$) & Nebulae$^a$ & Regions$^b$
& $t^2 \neq 0.000$ & $t^2 =0.000$\\
\hline 

O/H  & $-0.058$ & $-0.055\pm0.015$ & $-0.049$ & $-0.032$ \\
Ne/H & $-0.036$ & $-0.062\pm0.020$ & $-0.045$ & $+0.031$ \\
S/H  & $-0.077$ & $-0.062\pm0.020$ & $-0.055$ & $+0.006$ \\
Cl/H &    ...   & $-0.031\pm0.030$ & $-0.031$ & $+0.021$ \\
Ar/H & $-0.051$ & $-0.044\pm0.030$ & $-0.044$ & $-0.004$ \\

\hline
\multicolumn{5}{l}{$^a$ \scite{mac99}.}\\
\multicolumn{5}{l}{$^b$ \scite{est99a}.}\\
\multicolumn{5}{l}{$^c$ \scite{est99b}.}\\
\multicolumn{5}{l}{$^d$ \scite{est98}.}\\
\end{tabular}
\end{center}
\end{table*}

There are only three \ion{H}{2} regions with good determinations of $t^2$:
Orion, M8, and M17; in Table 2 we present for these \ion{H}{2} regions two
sets of gradients derived under the assumption of $t^2 = 0.000$ and $t^2 \neq
0.000$. The abundances derived under the assumption of $t^2 \neq 0.000$
indicate the presence of moderate gradients, while for those derived under the
assumption of $t^2 = 0.000$ the gradients disappear. This result supports the
contention that temperature fluctuations exist inside gaseous nebulae.

\section{Evolution of C/O with Time or with O/H in the Solar Vicinity}

\scite{car02} has constructed models of the chemical evolution of the Galaxy
based on observational yields of carbon derived from recombination lines and
from forbidden lines of planetary nebulae, she finds that the models that use
the yields based on permitted lines agree better with the observational
constrains provided by H~{\sc{ii}} regions and stars of the solar vicinity,
than the models based on the yields derived from forbidden lines. This result
is also indicative of the presence of temperature fluctuations inside gaseous
nebulae.

\section{$\Delta Y/ \Delta Z$ and $\Delta Y / \Delta O$ in our Galaxy, M101
and M33}

In Table 3 we present the following values in mass units: helium ($Y$), oxygen
($O$), and heavy elements ($Z$) for M17 computed in this paper. The helium
abundance was recomputed from the line intensities presented by \scite{est99a},
the He recombination coefficients computed by \scite{ben99}, the helium
recombination coefficients computed by \scite{sto95}, and the optical depth
effects computed by \scite{rob68}. The results presented in Table 3 are an
average for regions M17-3 and M17-14. The self consistent method (see Peimbert,
Peimbert, \& Luridiana 2002) based on the $\lambda\lambda$ 3819, 3889, 4026,
4387, 4471, 4922, 5876, 6678, 7065, and 7281 lines yielded $N_e = 800 {\rm
cm}^{-3}$ and $\tau(3889) = 7.5$ for M17-3, and $N_e =500 {\rm cm}^{-3}$ and
$\tau(3889) = 4.5$ for M17-14. The helium values are slightly smaller than
those presented by \scite{est99a}.

\begin{table*}[htb]
\begin{center}
\caption{Abundances of Helium, Oxygen and Heavy Elements by Mass}
\label{abundances}
\begin{tabular}{l@{\hspace{24pt}}c@{\hspace{12pt}}c
@{\hspace{12pt}}c@{\hspace{12pt}}c@{\hspace{12pt}}c@{\hspace{12pt}}c}
\hline
 & $Y$ & $Y$ & $O$ & $O$ & $Z$ & $Z$ \\
Object & $(t^2 = 0.000)$ & $(t^2\neq0.000)$
& $(t^2 = 0.000)$ & $(t^2\neq0.000)$
& $(t^2 = 0.000)$ & $(t^2\neq0.000)$ \\
\hline 
M17      & 0.2766 & 0.2677 & 0.00440 & 0.00849 & 0.0135 & 0.0201 \\
NGC 604  & 0.2708 & 0.2641 & 0.00427 & 0.00643 & 0.0096 & 0.0129 \\
NGC 5461 & 0.2706 & 0.2612 & 0.00501 & 0.00896 & 0.0113 & 0.0172 \\
\hline
\end{tabular}
\end{center}
\end{table*}

Also in Table 3 we present the abundance values for NGC 5461 in M101, and 
for NGC 604 in M33 derived by \scite{est02y}.

In Table 4 we present the $\Delta Y/\Delta O$ and the $\Delta Y/ \Delta Z$
values derived from a pregalactic helium abundance of $Y_p({\rm{+Hc}}, t^2 \neq
0.000) = 0.2384 \pm 0.0025$ and $Y_p({\rm{+Hc}}, t^2 = 0.000) = 0.2475 \pm
0.0025$ (Peimbert, Peimbert, \& Luridiana 2002) and the abundance values
presented in Table 3, where +Hc indicates that the collisional contribution to
the hydrogen Balmer lines has been taken into account.

\begin{table*}[htb]
\begin{center}
\caption{Values of $\Delta Y/\Delta O$ and $\Delta Y/\Delta Z$}
\label{Y/O-Z/O}
\begin{tabular}
{c@{\hspace{36pt}}c@{\hspace{18pt}}c@{\hspace{18pt}}c@{\hspace{18pt}}c}
\hline
 & \multicolumn{2}{c}{$\Delta Y/\Delta O$}
& \multicolumn{2}{c}{$\Delta Y/\Delta Z$} \\
\cline{2-3} \cline{4-5}\\[-9pt]
Object & $(t^2 = 0.000)^a$ & $(t^2\neq0.000)^b$
& $(t^2 = 0.000)^a$ & $(t^2\neq0.000)^b$ \\
\hline 
M17      & $6.60\pm1.25$ & $3.45\pm0.65$ & $2.15\pm0.45$ & $1.45\pm0.30$ \\
NGC 604  & $5.45\pm1.40$ & $4.00\pm1.00$ & $2.40\pm0.60$ & $2.00\pm0.50$ \\
NGC 5461 & $4.60\pm1.40$ & $2.55\pm0.80$ & $2.05\pm0.60$ & $1.35\pm0.35$ \\
\hline
\multicolumn{5}{l}{$^a$ $Y_p$(+Hc)=0.2475.} \\
\multicolumn{5}{l}{$^b$ $Y_p$(+Hc)=0.2384.} \\
\end{tabular}
\end{center}
\end{table*}

For M17 the difference between the values presented in 
Table 4 and those presented by \scite{est99a}, in addition to the
difference in the $Y$ values mentioned in the previous paragraph,  is due to the
adoption of two different pregalactic helium abundances, one for the
$t^2 = 0.000$ values and another for the $t^2 \neq 0.000$ values, while
\scite{est99a} used the same $Y_p$ for both types of determinations.

Based on their two-infall model for the chemical evolution of the Galaxy
\scite{chi97} find $\Delta Y/\Delta O = 3.15$ for the solar
vicinity. \scite{cop97} derives values of $\Delta Y/\Delta O$ in the 2.4 to 3.4
range. \scite{car00} computed chemical evolution models for the Galactic disk,
under an inside-out formation scenario, based on different combinations of
seven sets of stellar yields by different authors; the $\Delta Y/\Delta O$
spread predicted by her models is in the 2.9 to 4.6 range for the
Galactocentric distance of M17 (5.9~kpc), the spread is only due to the use of
different stellar yields. The $\Delta Y/\Delta O$ values predicted by the
models are in better agreement with the observations derived under $t^2 \neq
0.000$ the assumption than those derived under the $t^2 = 0.000$ assumption.

Similarly the models of chemical evolution of the Galaxy mentioned in the
previous paragraph predict $\Delta Y/\Delta Z$ values in the 1--2 range.  Again
the values derived from observations under the assumption that $t^2 \neq 0.000$
are in better agreement with the models than those derived under the assumption
that $t^2 = 0.000$.

\section{He/H Abundances in Planetary Nebulae and Giant Extragalactic
\ion{H}{2} Regions}

Each helium recombination line has a different dependence on the density and
the temperature, therefore from the helium line intensity ratios it is possible
to determine the density and the temperature in gaseous nebulae. \scite{pen95}
and \scite{pei95} determined $T$(\ion{He}{2}) values considerably smaller than
the $T$(\ion{O}{3}) values for a group of planetary nebulae. The extreme object
was Hu 1-2, where $T$(\ion{O}{3}) amounts to $19000 \pm 500$K, while
$T$(\ion{He}{2}) derived from the $\lambda\lambda$ 4472, 5876, and 6678 lines
amounts to $12500 \pm 500$K.

From extragalactic \ion{H}{2} regions Peimbert, Peimbert, \& Ruiz (2000) and
Peimbert, Peimbert, \& Luridiana (2002) also find that $T$(\ion{H}{2}) is also
smaller than $T$(\ion{O}{3}). This difference is paramount for the
determination of the primordial helium abundance because the use of
$T$(\ion{O}{3}) instead of $T$(\ion{He}{2}) leads to an overestimate of $Y_p$
of about 0.008, a small amount but significant for testing the standard
big-bang model and possible deviations from it.

\bigskip

It is a pleasure to acknowledge information prior to publication and fruitful
discussions with Leticia Carigi, C\'esar Esteban, Valentina Luridiana, Miriam
Pe\~na and Silvia Torres-Peimbert.


\end{document}